\documentclass[doublecol]{epl2} 
\usepackage{color}

\title{Weak Magnetic Order in High-$T_c$ Superconductors Produced by
Spontaneous Josephson Currents}
\shorttitle{Spontaneous Ferromagnetism in High-$T_c$ Superconductors} 

\author{E. V. L de Mello\inst{1,2} \and David M\"ockli\inst{2}}
\shortauthor{E. V. L de Mello \etal}

\institute{                    
  \inst{1} Department of Physics, Simon Fraser University - Burnaby, British Columbia, Canada V5A 1S6\\
  \inst{2} Instituto de F\'{\i}sica, Universidade Federal Fluminense - Niter\'oi, RJ 24210-340, Brazil
}

\abstract{
We develop a model for high-$T_c$ superconductors based on 
an electronic phase separation where 
low and high density domains are formed. At low temperatures this system
may act as a granular superconductor forming an array of Josephson
junctions. Cuprates are also known to have low superfluid densities
and strong correlation effects. Both characteristics activate a negative
Josephson coupling due to frustration that leads to spontaneous currents
responsible for the weak ferromagnetic order. This original approach reproduces
the observed onset of spontaneous magnetic signal and its dependence on
the doping level.}

\begin{document}

\maketitle

\section{Introduction}

Understanding of the normal state "pseudogap" phase of high-temperature
superconductors remains one of the major puzzles of condensed
matter physics. The reason abides in unusual properties and  partial 
gapping of the low energy density of states below a certain 
temperature $T^*$\cite{TS,PLee,NPK}, taken as the boundary of the
pseudogap phase.
The lack of an accepted theory hints the nanoscale
complexity and the intrinsically inhomogeneous electronic 
structures\cite{Muller,Dagotto}, which might vary according to the 
specific family of cuprates.

Among the many anomalies of the pseudogap phase\cite{TS,PLee,NPK},
the weak ferromagnetic signal on  YBa$_2$Cu$_3$O$_{6+x}$ by 
zero-field muon spin relaxation\cite{SSonier} and by polar Kerr effect\cite{Xia}
(PKE) requires an explanation. Both measure
a hole concentration ($p$) dependent signal that starts jointly
with $T^*(p)$ in agreement with
many different experiments\cite{TS,PLee,NPK}.
The values of $T^*(p)$ are well above $T_c(p)$ in the underdoped region
and drops rapidly with increasing $p$, becoming comparable with
$T_c$ near the optimally doped concentration $p=0.16$.

In this letter we address this long standing problem by an entirely
original approach. We assume an electronic phase separation with
the charges segregated in domains (grains) of
low and high densities separated 
by a potential barrier\cite{CH,Mello09,EPL12a}.
This charge segregation reduces the kinetic energies enhancing  
the possibility of superconducting amplitude formation in the 
isolated grains as a granular superconductor. This
local superconductivity in domains is possibly similar to the
case in which Bi-clusters of 2.5 to 40 nm becoming superconducting,
in contrast with the non superconducting bulk Bi\cite{Micklitz}.
The free energy barriers between the domains produces charge
confinement, as in the Bi-cluster case, and acts as weak links. 
We model this system as an array of 
Josephson junctions. Low local charge densities yield
strong phase fluctuations \cite{Em} that combined with strong correlation effects
gives rise to frustration\cite{SK}.  Our breaking new idea proposes 
that these frustration effects are the origin of the weak ferromagnetic
signal in YBCO\cite{SSonier,Xia}, which is a generalization to single crystals 
of the method used to explain the paramagnetic response of 
granular cuprates\cite{SR}.

\begin{figure}[ht]
    \begin{center}
     \centerline{\includegraphics[width=6.5cm]{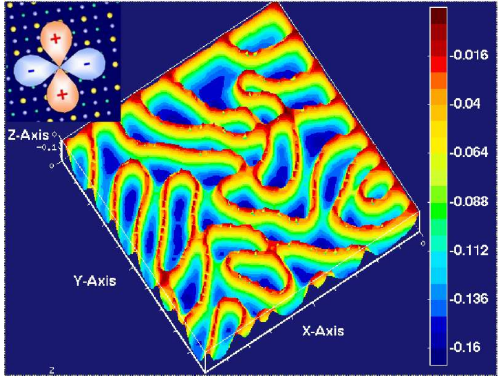}}
\caption{ (color online) Typical simulation of the Ginzburg-Landau free energy 
potential ${\it V_{GL}}(u,T)$ on a $105 \times 105$ square lattice. 
The grain's diameter spans a few nanometers. At low 
temperatures the charges are attracted to the grains, similarly to
a granular superconductor. We show the d-wave amplitude on the top left. 
}
\label{MapVu} 
\end{center}
\end{figure}

Many reviews\cite{Muller, Dagotto} have discussed the existence of electronic 
inhomogeneities by different experiments on several samples.
The disorder of the charge
in cuprates can be treated by the Cahn-Hilliard differential equation
that describes the formation of patches or grains.
The charges trapped in these grains lose kinetic energy, form bound states
and Cooper pairs, give origin to the pseudogap and superconducting
phases\cite{Mello09,EPL12a}.
With this, we illustrate a typical Ginzburg-Landau 
free energy simulation\cite{CH}
on a 105$\times$105 square lattice in Fig.(\ref{MapVu}).

Earlier experiments on granular 
cuprate superconductors showed anomalous
magnetic properties attributed to frustration due to
properties of the Josephson junctions\cite{SR}. In this case, the frustration
effects  arise due to the crystal's axis mis-orientation
of the grains with superconducting d-wave order parameter; this 
gives a negative contribution to the Josephson-junction energy. 
The current is negative with a phase shift $\pi$ -a $\pi$ junction. Sigrist
and Rice\cite{SR} demonstrated that a loop of current with an odd number of $\pi$
junctions is frustrated, i.e., there is no way to minimize the free energy
of all junctions in the loop and a spontaneous current may arise if the 
coupling is sufficiently strong.

\begin{figure}[ht]
\begin{center}
     \centerline{\includegraphics[width=6.0cm,angle=0]{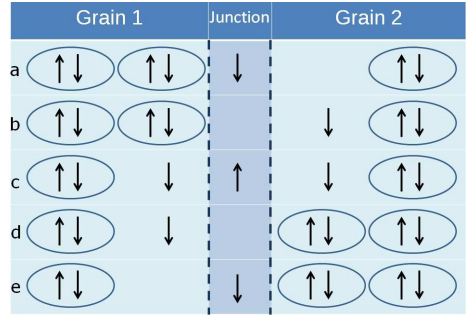}}
\caption{A schematic example of a sequence of intermediate states for 
a Cooper pair tunneling across a wall which contains a spin up and 
leads to a negative coupling along the junction. 
Between steps b and c it is necessary to permute the two 
electrons to be in the canonical order of a pair, and this exchange
is responsible for the negative sign\cite{SK}. }
\label{Spivak}
 \end{center}
\end{figure}

However, we cannot talk of mis-alignment of the crystal axis between 
different grains on a single crystal. In such system, negative Josephson
coupling is possible due to the large fluctuation in the local density
and correlation effects. This  was demonstrated in detail by 
Spivak and Kivelson\cite{SK} using perturbation theory up to $4^{th}$-order 
in a model where correlations effects 
produce a negative Josephson coupling across a junction. For completeness
we illustrate one of the situations described in their paper where a
Cooper pair tunnels through a
barrier which contains a spin up as shown in the Fig.(\ref{Spivak}).

It is important to remark that the physics of d-wave Josephson junctions presents
many different and interesting aspects, depending on the type of superconductors,
order parameter symmetry and geometry of the interface \cite{Gol}. 
The d-wave pairing order parameter is very sensitive to inhomogeneities and 
interfaces. Quasiparticle scattering at interface distorts the order parameter
and influences the Josephson effect as well as the quasiparticle tunneling
current \cite{Barash96}. This phenomenon can generate midgap states, i.e.,
surface or interface states with zero energy relative to the Fermi energy \cite{Hu}
or Andreev zero-energy. The midgap states enhance the Josephson current near
$T=0$ and are observed through the zero bias conductance peak \cite{Hu, Tanaka, Fogelstrom}.  
These states are robust phenomena
and represent a crossover from a traditional Josephson junction ($0$-junction)
to a $\pi$-junction, or a phase with broken time reversal symmetry.

On the other hand, the very high temperatures up to 200K
of the signal measured by Xia et. al. \cite{Xia} shows that its origin is not likely to be
due to these interface states. Furthermore, 
the intrinsic inhomogeneous nanometer domains have small and randomly
oriented interface, in opposition to those specially prepared junctions \cite{Hu, Tanaka, Fogelstrom},
and consequently it is difficult to generate midgap states. Thus, here we explore the idea that the origin of the $\pi$-junctions and the spontaneous 
magnetization in YBCO is due the very large phase oscillations due to very
small local density of Cooper pairs \cite{Em}.

\section{Calculations}

The Cooper pairs tunneling through a localized
state in the junction between two grains exchange order; this generates 
the negative sign -a characteristic  of 
a $\pi$ junction\cite{SK}. 
Since the electronic phase separation with the granular structure provokes 
strong charge fluctuations, -this- and not the
mis-orientation of the grains, is the mechanism originating $\pi$ 
junctions throughout the system.

Following the procedure of Sigrist and Rice\cite{SR},  
the properties of the whole loop of an odd number of $\pi$ junctions
is given at one weak link. Assuming that
the current $I$, which flows in the loop, is small compared to the 
critical current of the grains, we can write\cite{SR}

\begin{equation}
F(I,\Delta\phi)= \frac{LI^2}{2c^2}-\frac{\Phi_0I_c}{2\pi c}\cos\left(\Delta\phi+\alpha \right),
\label{FEloop}
\end{equation}

where $\Delta\phi$ is the phase shift across the junction, $L$ denotes the
self-inductance of the loop, $I_c$ the critical current
through the junction, and $\Phi_0$ is the flux quantum $hc/2e$. The 
additional phase shift $\alpha$ is $\pi$ or 0  whether the loop is frustrated
or not. Assuming also that the current flows through one point,
this leads to

\begin{equation}
 \Delta\phi =2\pi n-2\pi\frac{\Phi}{\Phi_0}=2\pi n-\frac{2\pi}{\Phi_0}\left(\Phi_{ex}-\frac{LI}{c}\right),
 \label{Fase}
\end{equation}
where $n$ is an integer, $\Phi$ is the total flux threading the loop with 
three types of contributions;
the external field, the flux from other loops called $\Phi_{ex}$ 
and the current $I$. With this $\Delta\phi$ Eq.{\ref{FEloop} becomes\cite{SR}
\begin{equation}
 F(I,\Phi_{ex})=\frac{LI^2}{2c^2}-\frac{\Phi_0I_c}{2\pi c}\cos\left[\frac{2\pi}{\Phi_0}\left(\Phi_{ex}+
 \frac{LI}{c}\right)+\alpha\right].
\end{equation}

By minimizing $F$ with respect to $I$ for a given $\Phi_{ex}$, we relate
$I$ and $\Phi_{ex}$. At zero external field we are interested in the
case where $\Phi_{ex}\approx 0$ because
the contributions from the other loops should be small. Then, for  $\alpha=\pi$,
a spontaneous flows around the loop if  $\gamma \equiv 2\pi LI_c/\Phi_0c$
is larger than 1 (Fig.(\ref{Isp}). This spontaneous current -due to the 
loop's frastration - generates the observed weak magnetic signal.

\begin{figure}[ht]
\begin{center}
     \centerline{\includegraphics[width=6.0cm,angle=0]{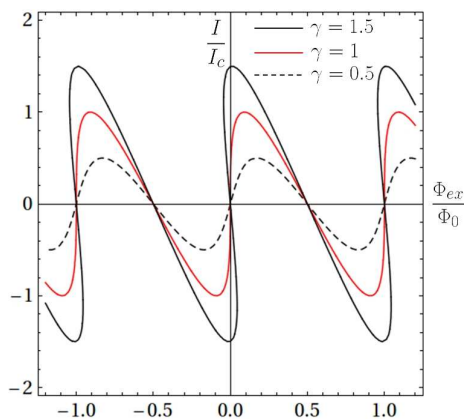}}
\caption{ The induced current in a frustrated loop for three values of the parameter
$\gamma$ as function of the external field. For the zero field case ($\Phi=0$),
it shows that for $\gamma >1$ a spontaneous current
may arise  while there is not any for $\gamma<1$. }
\label{Isp}
 \end{center}
\end{figure}

Thus, as the parameter $\gamma$ surpasses unity, a spontaneous current flows
around frustrated loops and local magnetic signals arise.
The first spontaneous currents generate a magnetic field that influences the 
direction of other currents and small local inhomogeneous ferromagnetic 
signals arise throughout the sample, similar to the zero field susceptibility
of granular superconductors near $T_c$\cite{Svedlindh} ($T^*$ in our case). 
Muon spin relaxation ($\mu$-SR)\cite{SSonier} and polar Kerr-effect\cite{Xia}
measured such magnetic signal on YBCO compounds.

To compare our proposal with these experimental results
we need to study the temperature and doping dependence of the
parameter $\gamma$, since it
controls the spontaneous currents. $\gamma$ needs to be larger than unity 
in order that a spontaneous current arises. On the other hand, $\gamma$ is
proportional to the critical current $I_c$, and this dependence  
was studied in detail for a d-wave superconductor by many 
authors\cite{Barash,Bruder}.

In particular Bruder et al\cite{Bruder} calculated the supercurrent tunnel 
matrix elements in second-order perturbation
theory for two d-wave superconductors (1 and 2) with superconducting
amplitude 
\begin{equation}
 \Delta_{d,1/2}(i,T,\phi)=\Delta(i,T)\cos\left[2(\phi-\phi_{1/2})\right],
\end{equation}
where $\phi$ is the azimuthal and $\phi_{1/2}$ is the mismatch angle 
which is here zero because the electronic domains are always aligned
with the crystal axis. The dominant contribution to the critical
current is from ``node to node'' tunneling and  the overall 
behavior is like an s-wave superconductor\cite{Bruder}. This result
is in agreement with the calculations of a Josephson junction 
with two different s-wave superconductors\cite{Anderson}, which yields
a supercurrent proportional to $\Delta_1\Delta_2/(\Delta_1+\Delta_2)$.

\begin{figure}[ht]
     \centerline{\includegraphics[width=5.5cm,angle=-90]{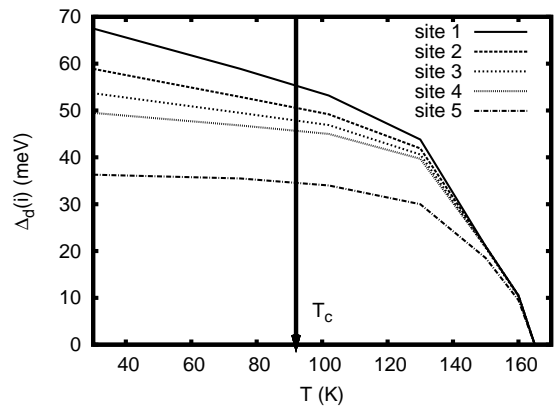}}
\caption{The temperature evolution of $\Delta_d(i)$ at five different locations 
of an average $p=0.15$ compound. The mean-field BdG calculations all vanish at the same 
temperature $T^*(0.15)\approx 165$K (above $T_c(0.15)=92$K). The low 
temperature values  between 30-66meV are in the energy
range of the Bi2212 LDOS gaps measured by STM.}
\label{DeltaiT} 
\end{figure}
 
Bogoliubov-deGennes calculations demonstrate that the values of the local 
superconducting amplitude also vary locally inside a system with local
electronic disorder\cite{PhysicaC2012,EPL12,Mello11}.
In Fig.(\ref{DeltaiT}) we show a typical result of these calculations and
plot the temperature evolution of the 
superconducting amplitude $\Delta_d(i) $ at five randomly chosen 
different locations or domains $i$.

\begin{figure}[ht]
\begin{center}
     \centerline{\includegraphics[width=5.5cm,angle=-90]{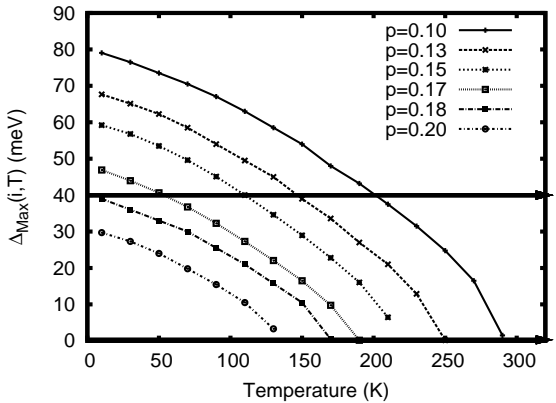}}
\caption{ The  temperature evolution of the maximum amplitude of the d-wave 
gap $\Delta(p,T)$ of six different values of average doping  $p$. 
The line at $\approx 40$meV shows the experimental onset of spontaneous
magnetization and the intersections yield the temperature onset for
different compounds.}
\label{IntDeltapT}
 \end{center}
\end{figure}

The disorder of the local gaps implies different 
values of the tunneling matrix between different grains. 
Consequently the values of $\gamma$  also
change at different junctions ($\gamma(i)$). The  values
of $\gamma(i)$ increases jointly with $\Delta(i,T)$ 
as the temperature decreases. Therefore the onset of spontaneous current occurs when
{\it the largest} $\Delta(i,T)$ reaches a critical value causing some 
junctions of a certain loop(s) to have $\gamma_{max}\ge 1$ 
at zero external field, as shown in Fig.(\ref{Isp}).

\section{Results}
It is very difficult to 
calculate theoretically the actual values of $\gamma(i)$ but we can obtain
some estimation from the experimental data. For instance, it is
reasonable that $\Delta(i,T)$ is the major
parameter that controls $\gamma(i)$ as the hole doping $p$ is varying.
The potential barrier between the grains is important but it is
also connected with the values of the superconducting amplitude.
Underdoped systems are in general more disordered than the overdoped ones, 
then it is possible that the number of loops is larger in this limit
increasing the strength of the ferromagnetic signal. Indeed the PKE signal
observed in underdoped samples are much stronger than those measured 
near optimally doped\cite{Sonier09}. Furthermore, below
$T_c$ the phase coherence diminish the phase oscillation
causing the signals measured below $T_c$ to be so small
that can be almost attributed to oscillation of the data\cite{SSonier,Xia}.

Since the magnetic signal vanishes near $p=0.18$,  we take the onset of $\Delta(i,T)$
that triggers the spontaneous current as the value shown in Fig.(\ref{IntDeltapT})
of $\Delta_{max}(p=0.18,T=0) \approx 40)$meV. In other words, a spontaneous
current can exist only above  $\Delta \approx 40$meV.

\begin{figure}[ht]
\begin{center}
     \centerline{\includegraphics[width=6.50cm]{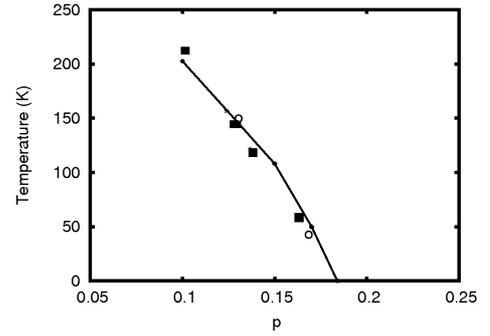}}
\caption{ The calculated  onset of spontaneous magnetization as derived from the
intersections in Fig.(\ref{IntDeltapT}) of the $\Delta_{max}(p,T)$
with the 40meV horizontal line together with the 
the experimental results of $\mu$-SR\cite{SSonier} (open circles) 
and polar Kerr-effect\cite{Xia}(black squares). 
The line connecting the theoretical points
are just a guide to the eyes.}
\label{Onset}
 \end{center}
\end{figure}

With this maximum value taken as the onset to have
$\gamma$ larger than 1, we can follow the temperature evolution of the 
maximum gap of each compound with the line drawn at  $\Delta_{max}(p=0.18,T=0)$
(Fig.(\ref{IntDeltapT})).
The results for 5 doping values  are shown in
Fig.(\ref{Onset}) together with the 
the experimental results of $\mu$-SR\cite{SSonier} (open circles) and 
of polar Kerr-effect\cite{Xia} (black squares).

\section{Conclusions}
In summary, we used the fact that cuprate superconductors have 
an intrinsic electronic disordered state where the charges are
segregated in  a few nanometers high and low density grains separated by 
thin potential walls. As in Bi-clusters\cite{Micklitz} this charge confinement
may be the origin of the local superconducting interaction.
We then calculate the superconducting properties by a BdG
method using a phenomenological two-body potential proportional to
the energy barriers or walls between the grains\cite{EPL12}.
The various distinct regions are coupled 
forming an array of Josephson junctions that promote the
resistivity transition at low temperatures. The phase-number quantum
fluctuations are large enough and strong correlation effects 
promote frustration or negative Josephson coupling\cite{SK} that leads to 
spontaneous currents and an overall magnetization. 

This approach connecting an electronic disordered state forming an array of
Josephson junctions at low temperatures with frustration effects
provides an interpretation to the intriguing spontaneous ferromagnetic 
phenomenon on cuprates, a long standing problem. This novel calculation 
shows the importance of the charge inhomogeneities
and leads to a pseudogap phase with isolated regions of non-vanishing 
superconducting amplitudes without long range order.

\section{Acknowledgments}
We gratefully acknowledge partial financial aid from Brazilian
agencies FAPERJ and CNPq.

\end{document}